\documentclass{interact}
\usepackage[utf8]{inputenc}
\usepackage{amsfonts}
\usepackage{enumitem}
\usepackage{parskip}
\usepackage{amsmath}
\usepackage{mathabx}
\usepackage{graphicx}
\usepackage{listings}
\usepackage{tikz}
\usepackage{etoolbox}
\usepackage{url}
\apptocmd{\sloppy}{\hbadness 10000\relax}{}{}
\usepackage{fancyhdr}
\usepackage{algorithm}
\usepackage[noend]{algpseudocode}
\usepackage{rotating} 
\usepackage{amsthm}
\usepackage{lmodern,textcomp}
\usepackage{chngcntr}
\counterwithin{figure}{section}
\counterwithin{table}{section}
\graphicspath{ {figures/} }
\usepackage{array}

\usepackage{natbib}
\bibpunct[, ]{(}{)}{;}{a}{}{,}

\theoremstyle{definition}
\newtheorem{definition}{Definition}[section]
 
\theoremstyle{remark}

\begin{document}

\title{Evidence and Behaviour of Support and Resistance Levels in Financial Time Series}

\author{
\name{K. Chung\textsuperscript{a}\thanks{CONTACT A. Bellotti. Email: anthony-graham.bellotti at nottingham dot edu dot cn} and A. Bellotti\textsuperscript{b}}
\affil{\textsuperscript{a}Department of Mathematics, Imperial College London, United Kingdom; \\
\textsuperscript{b}School of Computer Science, University of Nottingham, Ningbo, China}
}

\maketitle

%\pagebreak
%\tableofcontents

\begin{abstract}
This paper investigates the phenomenon of support and resistance levels (SR levels) in financial time series, which act as temporary price barriers that reverses price trends. We develop a heuristic discovery algorithm for this purpose, to discover and evaluate SR levels for intraday price series. Our simple approach discovers SR levels which are able to reverse price trends statistically significantly. Asset price entering SR levels with higher number of price bounces before are more likely to bounce on such SR levels again. We also show that the decay aspect of the discovered SR levels as decreasing probability of price bounce over time. We conclude SR levels are features in financial time series are not explained simply by AR(1) processes, stationary or otherwise; and that they contribute to the temporary predictability and stationarity of the investigated price series.
\end{abstract}

\section{Introduction}
Support and resistance levels (SR levels) are prices in a financial time series where it is believed that price trends are likely to stop and reverse \citep{osler}. SR levels are well known and frequently used within the retail trading circle as a form of technical analysis to inform trading decisions \citep{taylor}. Technical analysis is a class of methods which involves the analysis of historical data to inform trading decisions in the present. Traditionally it incorporates data such as price trends and trading activities, but modern advancements in data collection has opened potentially useful data sources such as twitter and news feed \citep{feed}. Unlike fundamental analysis in which the intrinsic value of a financial asset is evaluated in terms of economics and/or politics, technical analysis focuses solely on exploitable price patterns in the past.

The use of technical analysis for trading decisions can be dated as far back to the mid-19th century \citep{edwin}, when it was better known as tape reading. Ticker tapes were once in use to display current stock prices and volume, and tape reading refers to the analysis of the information on a ticker tape. Fast forward to the modern-day, retail traders (individual investors investing in retail platforms) mainly understand technical analysis to be the analysis of historic price series to predict future price movements. A common way of applying technical analysis is to perform calculations on a rolling window of historic prices to produce a numerical indicator \citep{rolling window1} \citep{rolling window2}. This indicator can then be used to inform trading decisions, for example if the trader should invest in a particular financial asset.

Technical analysis by nature go against the well established efficient market hypothesis (EMH) \citep{emh}. The EMH states that asset prices fully reflect all available information, and that as a direct consequence it is impossible to generate returns in excess of risk free returns on a consistent basis by analysing asset prices only. There is currently no consensus on this conundrum in the academic community, as both opposing sides have equally numerous and convincing academic evidence to support their claims. For example, a recent review \citep{ta evidence} on studies of technical analysis has found overwhelmingly positive results on its profitability. On the other hand, a paper \citep{emh evidence} defends the EMH by invalidating its critics and concludes that the stock markets are not as predictable as they claim.

The most important philosophical foundation as to why technical analysis can be profitable is that retail traders believe history repeats itself, in the form of market inefficiencies which can be repeatedly exploited. The prevalence of high frequency trading in the past decades \citep{hft rise}, which is mostly dependent on analysing order book information, is a testament to the efficacy of technical analysis. For this paper, we maintain a neutral stance on this subject and use methods from both sides to validate our findings. 

Verification of and optimization of the profit of technical indicators is in continuous research. An example of such an indicator is the Bollinger bands \citep{bollinger}, in which there were substantial research efforts \citep{boll1, boll2, boll3, boll4} since its publication. Despite interests in understanding technical indicators in general, research on SR levels are few and far between. We postulate that the difficulty of establishing a universal technical indicator due to its subjective nature could be the main reason for this. Although research on market micro-structure related to SR levels such as price barriers \citep{price barrier} and price clustering \citep{price clustering} can be found, the earliest account that verifies the predictive capability of SR levels is documented by Osler in 2000 \citep{osler},  who have stated that ``the specific conclusion that exchange rates tend to stop trending at support and resistance levels has no precedent in the academic literature''. 

From a market micro-structure perspective, the root cause of SR levels can be attributed to a large volume of limit orders stacked at certain price levels. Indeed, economists have concluded the existence of SR levels is due to stacked limit orders at certain price levels \citep{price clustering}. Limit orders are buy/sell orders of a financial asset which are executed only if pre-determined prices are met; whereas market orders are orders which are executed immediately at the current price. SR levels are results of an imbalance order flow at certain price levels \citep{osler2}. This order flow is an intricate interplay between limit orders and market orders.

Osler concluded that published SR levels by financial firms between 1996 to 1998 has statistically significant ability to predict trend interruptions \citep{osler}; and that their predictive abilities can last up to 5 business days after publishing. \cite{memory} have also investigated the bounce probability of SR levels discovered in the tick data for 9 stocks in the London Stock Exchange in 2002, and concluded that prices tend to re-bounce than cross price values which are determined to be SR levels. Our work centres around the evaluation framework of \cite{memory}, and contributes to the literature by proposing a heuristic sequential algorithm for SR level detection in Section \ref{background section}, investigates the temporal effect of the detected SR levels in Section \ref{temporal section} and if our findings occur naturally in a theoretical efficient market in Section \ref{AR section}. Section \ref{conclusion section} concludes this paper and discusses future work.

\section{Methodology}\label{background section}
\subsection{Definitions}
At the time of writing, no clear mathematical definition for SR levels can be found in the literature. Qualitative description of SR levels on the other hand are well established in technical analysis manuals and there are little differences in their definitions \citep{osler}. For example a standard technical analysis reference states that ``Support levels indicate the price where the majority of investors believe that prices will move higher, and resistance levels indicate the price at which a majority of investors feel prices will move lower.'' \citep{ta reference}. \cite{memory} evaluates the efficacy of SR levels using a probabilistic approach but also have a qualitative definition. As far as a general approach is concerned, recent minima and maxima are as good approximates of SR levels as any \citep{osler}.

These definitions, though form the basis of our discovery algorithm in the later sections, are unsatisfying mathematically and are not concrete enough to work with. To carry out our investigations in the subsequent sections, we propose the following definitions for support and resistance level in a discrete time framework:
\vspace{0.1cm}
\theoremstyle{definition}
\begin{definition}{Support Level}\label{support def}
is an interval $[a, b]\in\mathbb{R}$ such that if $x_t\in[a, b]$, then $p(x_{t+\delta}>b)>p(x_{t+\delta}<a)$ for all $\delta\in\{1,\dots,\omega\}$.
\end{definition}

\theoremstyle{definition}
\begin{definition}{Resistance Level}\label{resistance def}
an interval $[a, b]\in\mathbb{R}$ such that if $x_t\in[a, b]$, then $p(x_{t+\delta}>b)<p(x_{t+\delta}<a)$ for all $\delta\in\{1,\dots,\omega\}$.
\end{definition}

Our proposed definitions acknowledge the varying width of an SR level with an arbitrary interval $[a, b]$, as well as its temporary nature by specifying an arbitrary value $\omega$ for the length of its existence. The choice to specify an interval for an SR level is twofold: relaxation of strict price levels such as using SR zones instead of SR levels are popular among technical traders, and such relaxation allows more price bounces to be observed for any interval which helps in detecting SR levels. In section \ref{algo section}, we argue that the optimal choice of interval width for our discovery algorithm is determined by the average price increment, which is in turn related to the resolution of the price series. For very high frequency price series, or in a theoretical continuous context, choosing the interval width to be 0 may be suitable. However, our definitions do not directly address the supply and demand imbalance in the order book which is one of the root causes of SR levels \citep{osler2}, but is sufficient for our subsequent analysis where only the asset price series is concerned. 

\begin{figure}[!htbp]
\begin{center}
\includegraphics[width=13cm]{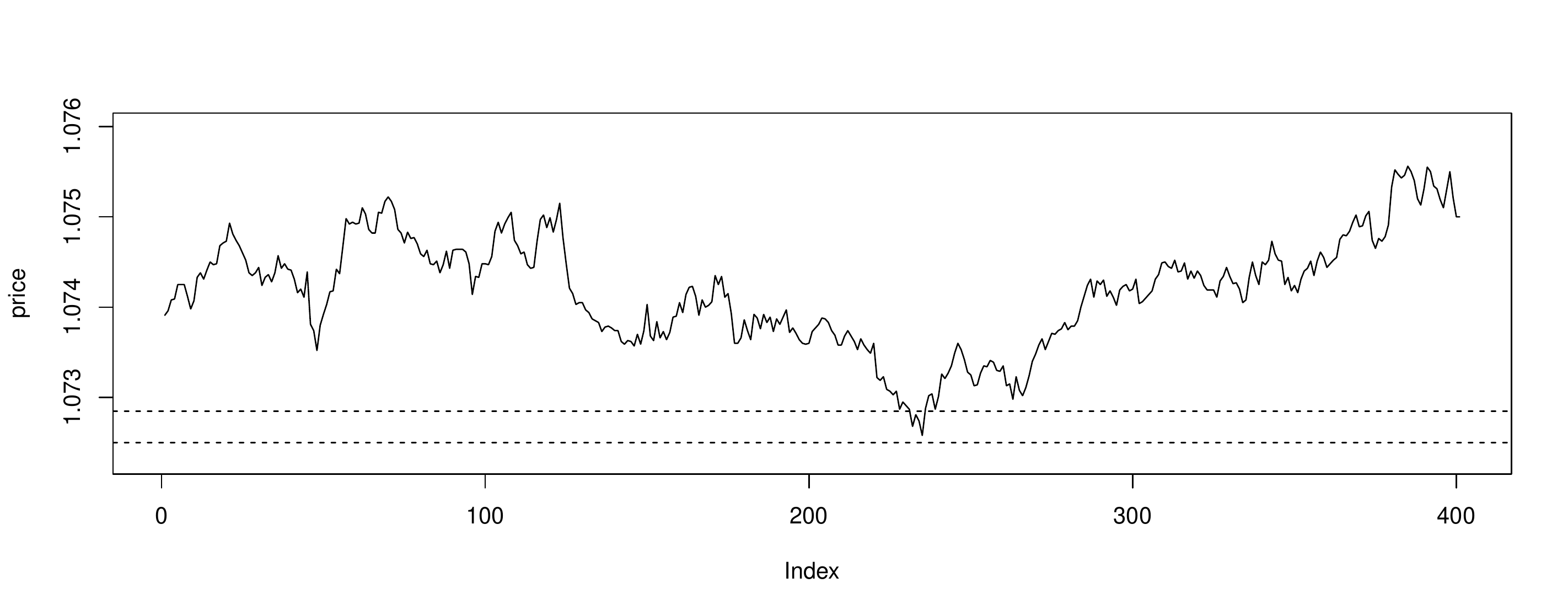}
\end{center}
\caption{Support level with a single bounce}\label{support example}
\end{figure}

\begin{figure}[!htbp]
\begin{center}
\includegraphics[width=13cm]{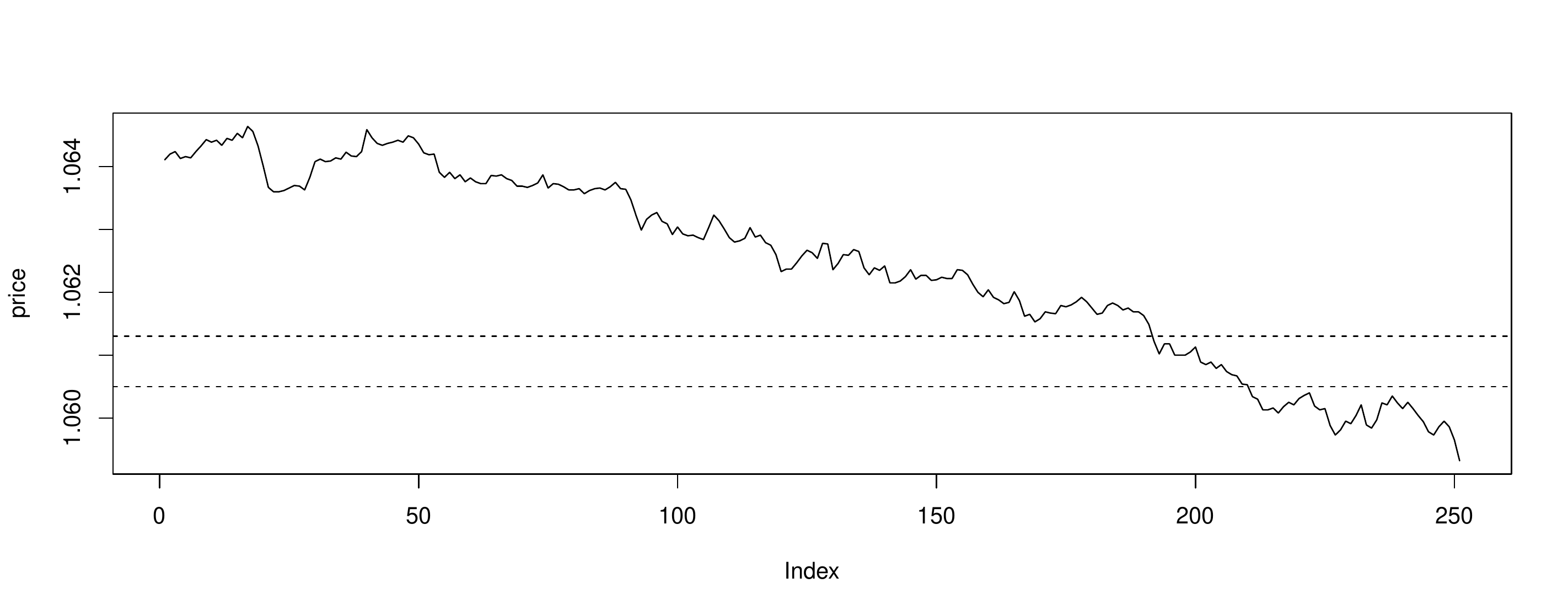}
\end{center}
\caption{Support level with penetration}\label{support penetration example}
\end{figure}

Next, we define the bounce and penetration event for an SR level. A bounce on a support (resistance) level is defined as the price entering the level's interval from its upper (lower) boundary and then exiting it through the upper (lower) boundary, without going below the lower (upper) boundary. We define penetration of a support (resistance) level as the price entering its region from the upper (lower) boundary and then exiting through the lower (upper) boundary. Figures \ref{support example} and \ref{support penetration example} display support levels with a price bounce and a price penetration respectively, with the dotted lines denoting the upper and lower boundary of the support level.

In order to evaluate the strength of an SR level following our proposed definitions, we define the probability of bouncing, which for a support level is
\begin{equation*}
p(b)=\frac{p(x_{t+\delta}>b)}{p(x_{t+\delta}<a)+p(x_{t+\delta}>b)},
\end{equation*}
and for a resistance level is
\begin{equation*}
p(b)=\frac{p(x_{t+\delta}<a)}{p(x_{t+\delta}<a)+p(x_{t+\delta}>b)},
\end{equation*}

where $x_t\in[a, b]$ and $\delta\in\{1,\dots,\omega\}$.  This probability of bouncing is conditional on price exiting the SR level, and is essential for our investigation in the subsequent sections. Note that the bounce probability defined here holds true for the particular price level from $t$ to $t+\delta$, for any $\delta\in\{1,\dots,\omega\}$, and that price exit events can happen at any time within this time interval. We also refer the strength of an SR level to be its $p(b)$ value; i.e. the higher the $p(b)$ value for an SR level, the stronger it is, therefore the better its ability to reverse price trends.

\subsection{Discovery \& Evaluation Algorithm}\label{algo section}
At the time of writing, no existing discovery algorithm of SR levels is documented in the academic literature. Osler evaluated SR levels published by financial institutions \citep{osler}, and \cite{memory} has not explicitly stated how their evaluated SR levels are discovered. The algorithm proposed in this section follows a rolling window approach commonly used in technical analysis \citep{rolling window}, and simply identifies local minima and local maxima as support and resistance levels respectively. We acknowledge that there are many methods for discovering SR levels in the trading circle and that therefore the discovery algorithm proposed here is one of all possible discovery methods. 

A rolling window approach entails that a pre-specified period of immediate historic price series being analysed by the algorithm at every time point. Given a rolling window with prices series $X_t=x_{\tau:\tau\in\{t-i, t-i+1,\dots,t\}}$, where $\tau$ is the time index and $i$ the rolling window length. We also refer to rolling windows with length $i$ minutes to be an $i$ minute lag window for the remainder of this paper. Following our definitions of an SR level earlier, we choose for a rolling window, a support level to be $\min X_t\pm\gamma$ and a resistance level to be $\max X_t\pm\gamma$, where $\gamma$ is the level width parameter. We follow the approach of Garzarelli et al. \citep{memory}, which accommodates different resolution of price series, by choosing $\gamma$ to be the average price increment of the entire price series defined as
\begin{equation}\label{sr width equation}
\Delta(x_t)=\frac{1}{T-1}\sum^T_{t=2}\vert x_t-x_{t-1}\vert,
\end{equation}
where $T$ is the last time index of the price series. We find that using this approach yields $p(b)=0.5$ consistently for random walk simulations which serves as a good benchmark, and that a higher $\gamma$ would inflate $p(b)$ while a lower $\gamma$ would deflate $p(b)$.

With the SR levels of a rolling window identified, we can then count the number of bounces there are for this pair of SR levels. In the case of a support level, its number of bounces $S_b$ can be determined by the number of crosses on the upper boundary of the support level $\bar{S}=\min X_t+\gamma$ of each price movement divided by 2; or in mathematical terms, the number of times $\bar{S}$ is between $x_{\tau-1}$ and $x_{\tau}$ for $\tau\in\{t-i, t-i+1,\dots,t\}$ divided by 2, where $i$ denote the rolling window length. Thus, the number of support bounces
\begin{equation*}
S_b=\Big\lfloor\frac{1}{2}\sum^{t-1}_{\tau=t-i}\mathbb{I}_{\{x_\tau\leq\bar{S}<x_{\tau+1}\}}+\mathbb{I}_{\{x_{\tau+1}\leq\bar{S}<x_{\tau}\}}\Big\rfloor,
\end{equation*}
where $\lfloor\:\rfloor$ denote the floor function, as single crosses are merely price entering support levels and do not constitute as bounces. Similarly, denoting $\underline{R}$ as the lower boundary of a resistance level, then the number of resistance bounces is
\begin{equation*}
R_b=\Big\lfloor\frac{1}{2}\sum^{t-1}_{\tau=t-i}\mathbb{I}_{\{x_\tau<\underline{R}\leq x_{\tau+1}\}}+\mathbb{I}_{\{x_{\tau+1}<\underline{R}\leq x_{\tau}\}}\Big\rfloor,
\end{equation*}

They can be calculated at each time step $t$ from the price series rolling window. The parameter $\gamma$ denotes the level width parameter.

This way of discovering SR levels is a heuristic approach and has some overly simplifying assumptions. It assumes SR levels do not last longer than the length of the rolling window, and therefore older price series does not contribute to the discovery of SR levels. We know this to be false as SR levels only become obsolete once the limit orders stacked at the price level are either executed or cancelled. It also assume there are only one pair of SR levels at any given time. This is also likely untrue as limit orders can be stacked with varying quantity at different price levels, leading to multiple SR levels with different strengths. Another caveat of this approach is that it strictly chooses SR levels to be the window maximum and minimum, effectively ignoring newly formed SR levels which are neither the maximum nor the minimum within the rolling window.

For the purpose of observing $p(b\vert b_{prev})$ described in the next section, the discovery procedure ceases to operate temporarily when price enters an SR level. This ensures that the discovered SR levels remain static, as the price could achieve a new minimum or maximum of the rolling window without penetrating the SR level. If the discovery procedure continues operating, a new minimum (maximum) would create a new lower (upper) boundary for the support (resistance) level, and thus erroneously reducing the probability of penetration. Once a bounce or penetration is determined, this is recorded along with the number of previous price bounces there has been on the SR level ($b_{prev}$), and the discovery procedure resumes process normally.

\subsection{Bounce Probability Bayesian Framework}\label{evaluation framework}
Retail traders believe that multiple price trend reversals on the same SR level indicate its strength. A framework to evaluate the conditional probability of a bounce $p(b\vert b_{prev})$ on an SR level given $b_{prev}$ previous bounces on the same SR level was developed for this purpose \citep{memory}. In this framework, $b$ denotes the binary variable where $b=1$ represents a bounce event and $b=0$ represents a penetration event, after asset price enters a predetermined SR level; and we impose no time limit for this SR level exit event to occur. This probability is also a method that measures the memory effect of financial time series \citep{universal memory}. To measure whether SR levels tend to interrupt price trends, we observe price behaviour within the SR levels. 

The framework \citep{memory} to estimate $p(b\vert b_{prev})$ continues as follows. Assuming the discovered SR levels are statistically identical and independent, and denoting

\begin{itemize}
\item $b_{prev}$ as the number of previous bounces of an SR level,
\item $n_{b_{prev}}$ as the total number of price bounce events given $b_{prev}$,
\item $k_{b_{prev}}$ as the total number of price penetration events given $b_{prev}$,
\item and $N_{b_{prev}}=n_{b_{prev}}+k_{b_{prev}}$ as the total number of times price exits given $b_{prev}$.
\end{itemize}

Then $p(b\vert b_{prev})$ can be modelled using the method of Bayesian inference. Assuming that $n_{b_{prev}}$ follows a Bernoulli process, as there are only two outcomes when the price enters an SR level - bounce or penetration, we can define the probability of success corresponding to a price bounce given the number of previous bounces of an SR level, as
\begin{equation*}
\pi=p(b\vert b_{prev}).
\end{equation*}

By assuming an uninformative uniform prior for $\pi$ such that
\begin{equation*}
Prior(\pi)=\mathcal{U}([0, 1]),
\end{equation*}
and modelling of observed $n_{b_{prev}}$ and $N_{b_{prev}}$ with a binomial likelihood
\begin{equation*}
l(n_{b_{prev}}\vert N_{b_{prev}},\pi)= \binom{N_{b_{prev}}}{n_{b_{prev}}}(\pi)^{n_{b_{prev}}}(1-\pi)^{N_{b_{prev}}-n_{b_{prev}}}.
\end{equation*}

The uniform prior is in fact a special case of the Beta distribution and can therefore be treated as a conjugate prior to a Beta posterior with respect to a binomial likelihood. As the posterior of $\pi$ can be written as 
\begin{equation*}
Posterior(\pi\vert N_{b_{prev}}, n_{b_{prev}})=Beta(\pi\vert n_{b_{prev}}+1;N_{b_{prev}}-n_{b_{prev}}+1)
\end{equation*}
The expectation and variance of $p(b\vert b_{prev})$ can be estimated using
\begin{equation*}
\begin{aligned}
\mathbb{E}[p(b\vert b_{prev})]&=\frac{n_{b_{prev}}+1}{N_{b_{prev}}+2}\\
Var[p(b\vert b_{prev}]&=\frac{(n_{b_{prev}}+1)(N_{b_{prev}}-n_{b_{prev}}+1)}{(N_{b_{prev}}+3)(N_{b_{prev}}+2)^2}
\end{aligned}
\end{equation*}

\section{Statistical Evidence}\label{existence}\label{memory section}
This section investigates the extend at which the hypothesised SR levels influence future price, which manifests as a memory effect. In particular, we investigate the relationship of the strength of an SR level and the number of previous price bounces it has. The result of a positively correlated relationship from analysing price series in 2018 in the foreign exchange market, London Stock Exchange and the commodity market, is consistent with the analysis by \cite{memory} on high frequency price series on 9 selected stocks in the London Stock Exchange in 2002. In addition, we obtain evidence for the temporal decay in this bounce probability and also show that simulated time series from AR1 processes are unable to reproduce similar results.

The minute intraday price series in 2018 for 3 financial assets are analysed: 1) the Euro-US dollar conversion rate (EURUSD) in the foreign exchange market, 2) the Lloyds Banking Group PLC (LLOY) equity in London Stock Exchange, and 3) the Brent Crude Oil (BRENT) commodity in the global commodity market. These financial assets are the most traded assets \citep{forex metrics} from their respective financial markets, with 372,607, 127,606 and 307,678 entries respectively after removing entries without trading activities. Note that stocks are only traded for 8 hours during trading days, and contributes to the vastly lower number of the LLOY price entries. The minute intraday price series required for analysis can be obtained for free from the following websites:
\begin{itemize}
\item Foreign exchange: http://www.HistData.com/
\item Stock/commodity: http://www.livecharts.co.uk/
\end{itemize}

\subsection{Estimated Bounce Probability}\label{existence results section}
The estimated $p(b\vert b_{prev})$ values evaluated by applying the discovery and evaluation algorithms described in Section \ref{algo section} are presented here. In essence, we determine local minima and maxima to be local support and resistance levels, then count the number of bounces these levels have in the recent past, and finally observe if prices in the future bounce or penetrate these levels. This procedure is carried out throughout the entire price series and results are then aggregated using the Bayesian estimation framework described in Section \ref{evaluation framework}. 

Estimated $p(b\vert b_{prev})$ values represents the probability of price bouncing off an SR level again given the number of previous bounces, and estimated values other than 0.5 suggests predictability in the price series. The higher this estimate is, the more likely price is going to bounce again on the discovered SR level. Since we are investigating the ability of SR level to reverse price trends, we expect the estimated $p(b\vert b_{prev})$ values to be higher than the baseline 0.5. To demonstrate this price reversal ability as a memory effect in the price series, we also estimate $p(b\vert b_{prev})$ for a shuffled returns price series for comparison. The confidence intervals in this section are constructed using a width of 1 standard deviation on each side which only represents a 68\% confidence interval. This is done to improve the readability of the charts due to some large standard deviations estimations. Doubling the width of the confidence intervals in the produced charts gives 95\% confidence intervals and may serve as a better comparison of the estimated $p(b\vert b_{prev})$ values between the original price series and the shuffled return price series.

\begin{figure}[!htbp]
\begin{center}
\includegraphics[width=13cm]{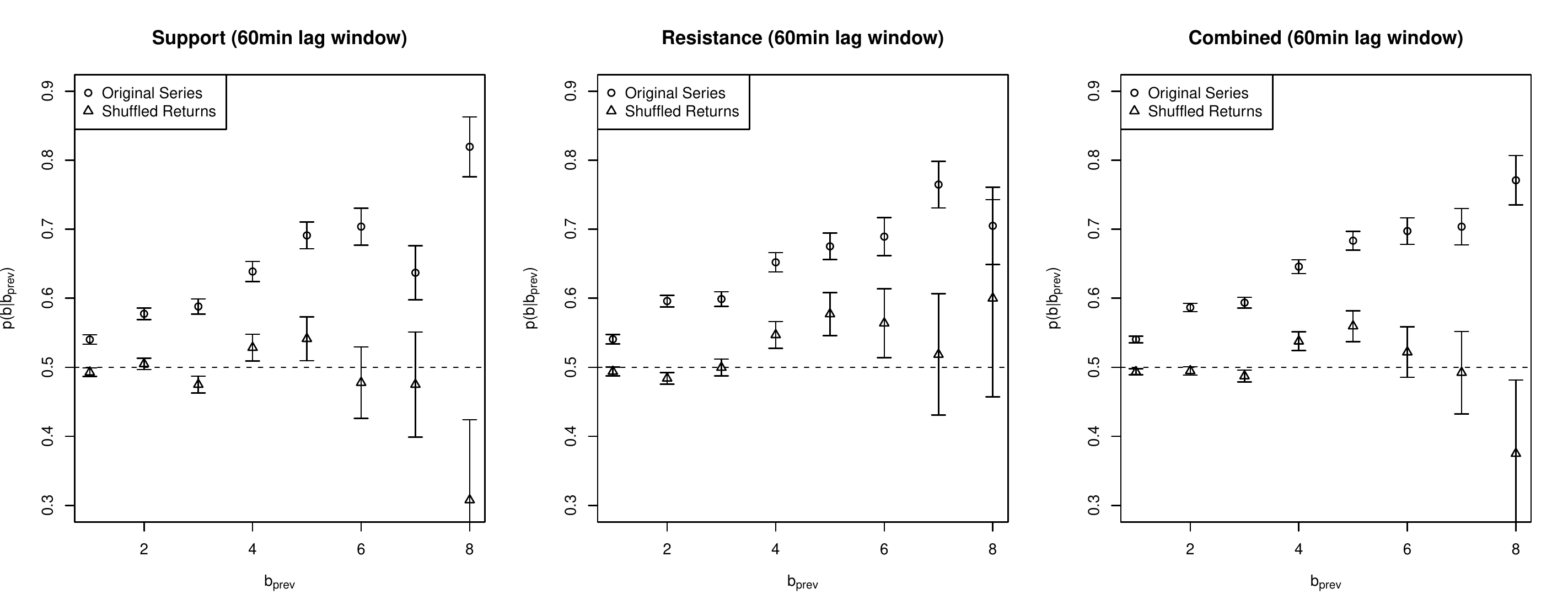}
\end{center}
\caption{\textbf{Estimated $p(b\vert b_{prev})$ for EURUSD (60-minute lag window).} The confidence intervals are constructed using a width of 1 standard deviation on each side.}\label{eurusdSR}
\end{figure}

Figure \ref{eurusdSR} displays the estimated $p(b\vert b_{prev})$, in which the results for the shuffled returns series is superimposed with the results for the original prices for comparison. Directing our attention to the estimated $p(b\vert b_{prev})$ for the original series, it is seen that $b_{prev}$ is positively correlated with $p(b\vert b_{prev})$. This result coincides with the stylised fact of multiple price trend reversals of an SR level confirms its strength; and provides evidence for the self-reinforcement hypothesis in the behavioural finance perspective \citep{prophecy} where traders collectively bet on price trend reversals at SR levels with a high number of previous bounces. Furthermore, the estimated $p(b\vert b_{prev})$ is higher for the original price series than that of the shuffled returns series. There is also no visually discernible difference in price behaviour between support and resistance levels. This comparison with the shuffled returns series provides evidence for a memory effect in the price series, at least in the short term. Lastly, having $b_{prev}=8$ in the shuffled returns series is an occurrence so rare that there are only 16 recorded such SR level entry events. This small sample size has also affected the $p(b\vert b_{prev})$ estimation and its standard deviation. Due to this, it might become difficult to determine statistically if $p(b\vert b_{prev}=8)$ is indeed higher in the original price series, but it is logical to assume $p(b\vert b_{prev}=8)$ converges to 0.5 for a perfect random walk which has no memory.

\begin{figure}[!htbp]
\begin{center}
\includegraphics[width=13cm]{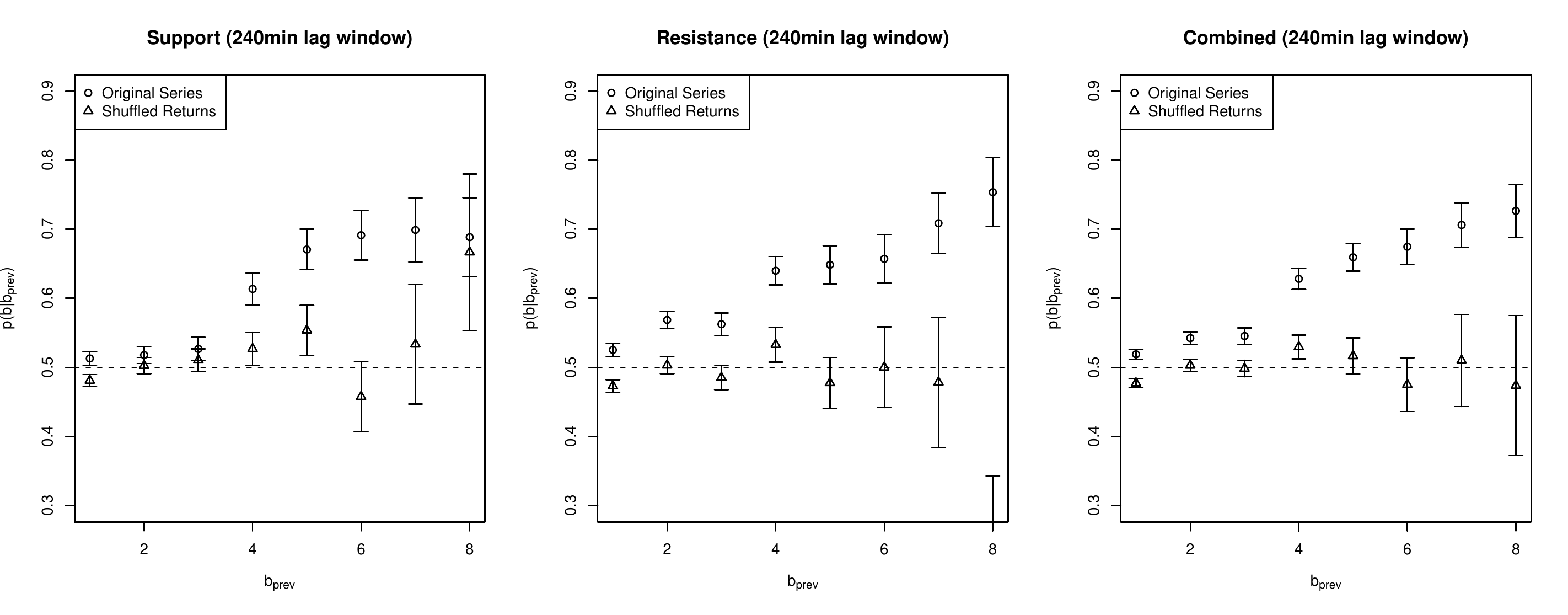}
\end{center}
\caption{\textbf{Estimated $p(b\vert b_{prev})$ for EURUSD (240-minute lag window).} The confidence intervals are constructed using a width of 1 standard deviation on each side.}\label{eurusdSR240}
\end{figure}

We perform the same procedure on a longer lag window to demonstrate SR levels exist also on other timeframes. Figure \ref{eurusdSR240} displays the results for the EURUSD 2018 price series but applying the discovery and evaluation algorithms using a lag window of 240 minutes. The results in general are similar to the results in the previous section for a lag window of 60 minutes: higher number of previous bounces predicts a higher probability of bouncing again. There is however, a decrease in the estimated $p(b\vert b_{prev})$, which confirms the temporary nature of SR levels, i.e. the strength of SR levels decreases with time. There is also a decrease in the number of events recorded, up to half as many as the 60-minute lag window results; and is more prominent for lower $b_{prev}$ values. This is a consequence of using a large lag window and that SR levels identified in a longer time frame simply has more opportunities for price bounces to occur, leading to less SR levels with a small number of previous bounces.

\begin{figure}[!htbp]
\begin{center}
\includegraphics[width=13cm]{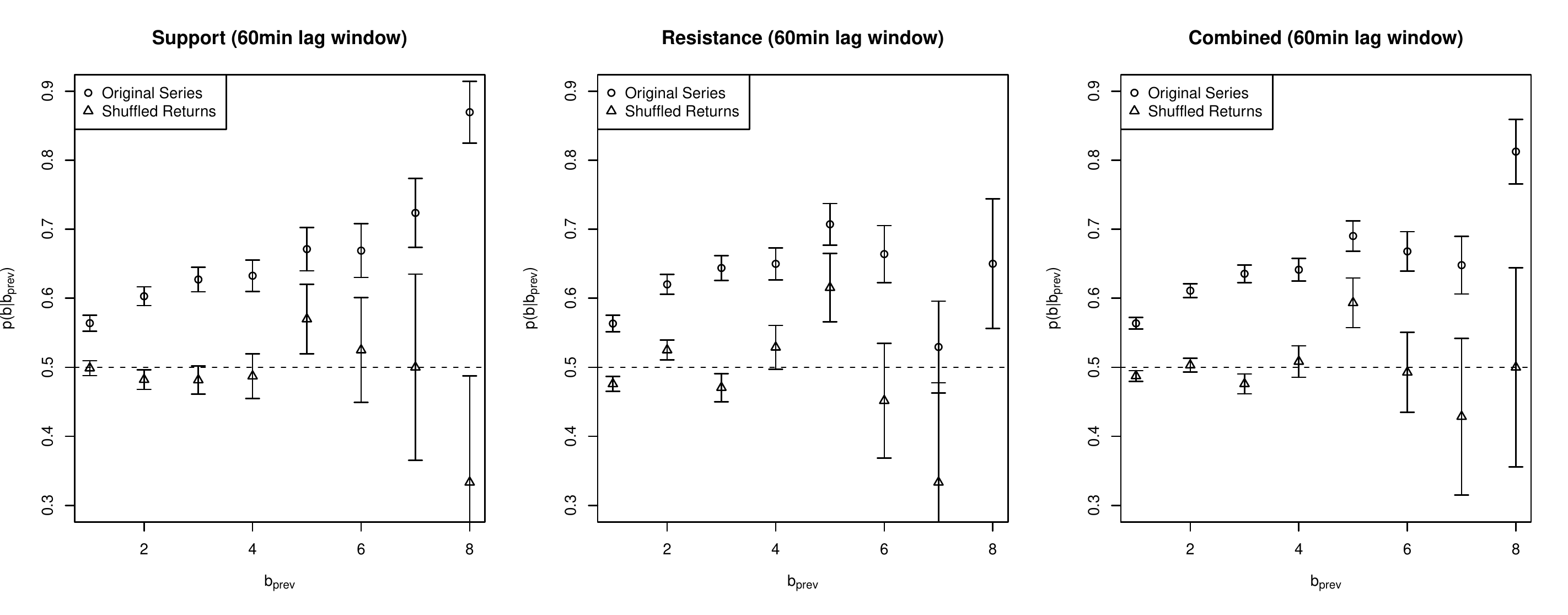}
\end{center}
\caption{\textbf{Estimated $p(b\vert b_{prev})$ for LLOY (60-minute lag window)}. The confidence intervals are constructed using a width of 1 standard deviation on each side.}\label{lloy60}
\end{figure}

\begin{figure}[!htbp]
\begin{center}
\includegraphics[width=13cm]{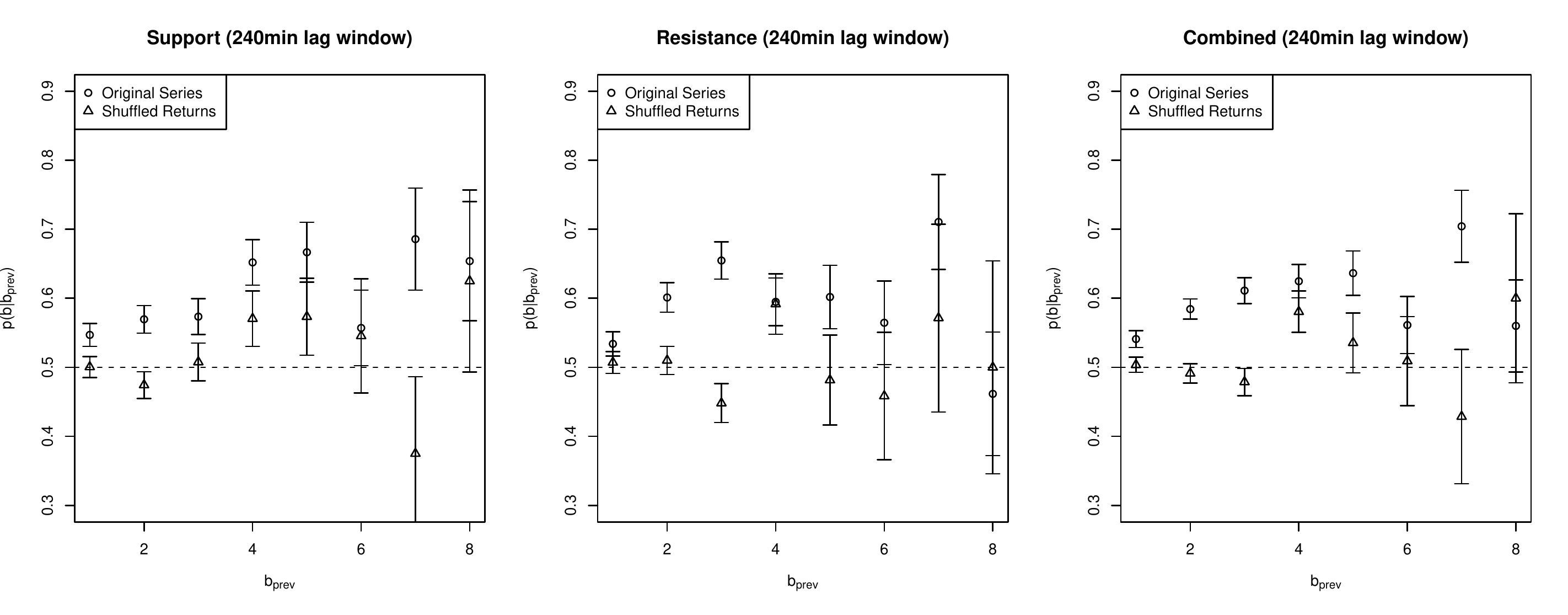}
\end{center}
\caption{\textbf{Estimated $p(b\vert b_{prev})$ for LLOY (240-minute lag window).} The confidence intervals are constructed using a width of 1 standard deviation on each side.}\label{lloy240}
\end{figure}

\begin{figure}[!htbp]
\begin{center}
\includegraphics[width=13cm]{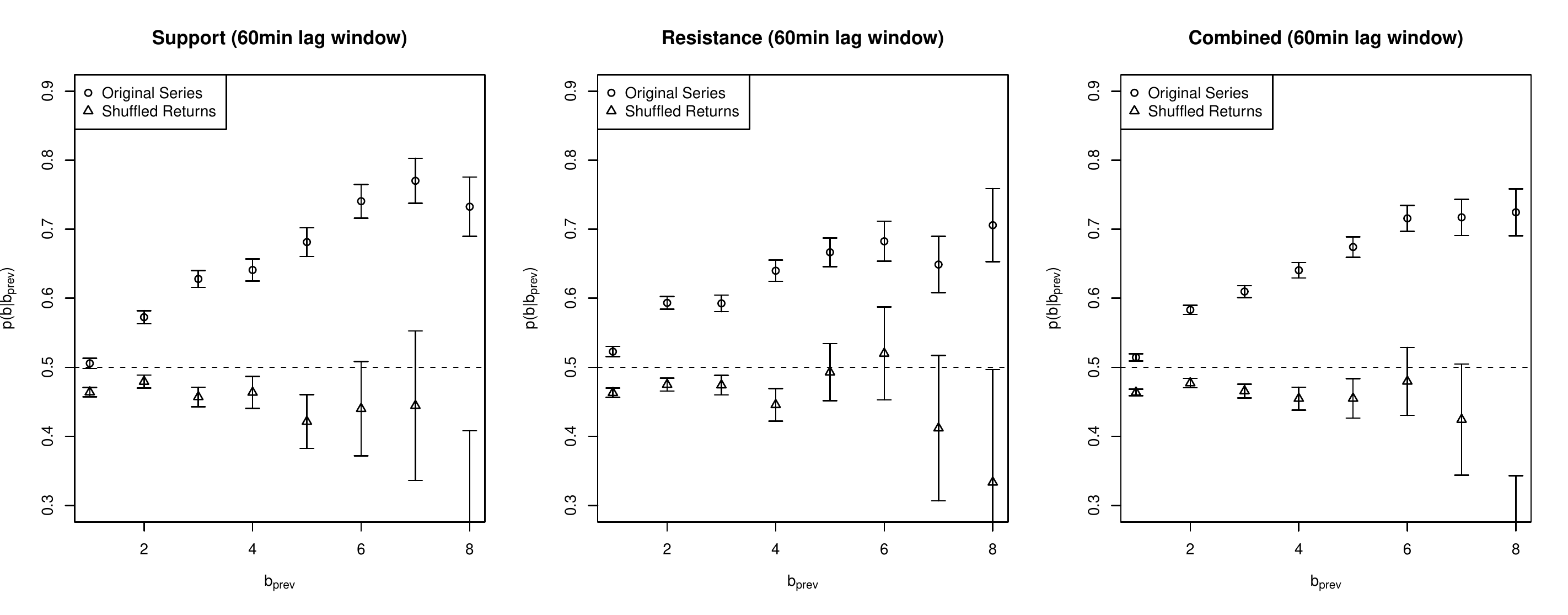}
\end{center}
\caption{\textbf{Estimated $p(b\vert b_{prev})$ for BRENT (60-minute lag window).} The confidence intervals are constructed using a width of 1 standard deviation on each side.}\label{brent60}
\end{figure}

\begin{figure}[!htbp]
\begin{center}
\includegraphics[width=13cm]{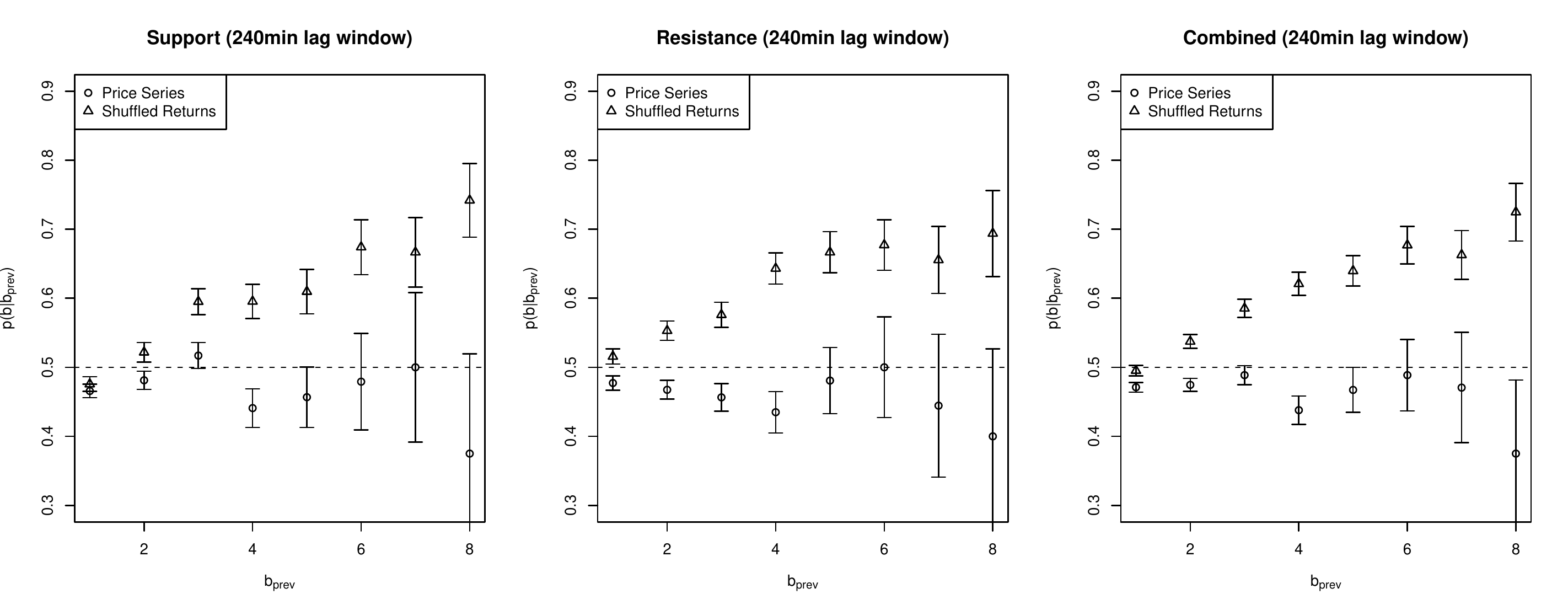}
\end{center}
\caption{\textbf{Estimated $p(b\vert b_{prev})$ for BRENT (240-minute lag window).} The confidence intervals are constructed using a width of 1 standard deviation on each side.}\label{brent240}
\end{figure}

Results for LLOY using a 60 and 240 minute lag window are displayed in Figure \ref{lloy60} and \ref{lloy240} respectively; the results for BRENT using a 60 and 240 minute lag window are displayed in Figure \ref{brent60} and \ref{brent240} respectively. The observed bounce behaviour is also similar for the LLOY and BRENT price series, and provides evidence for the universal existence of SR levels across asset classes. We again observe a positive correlation between an SR level's bounce probability and its number of previous bounces. As a side note, the LLOY price series has a length of approximately a third of the length of the EURUSD and BRENT series due to its market opening hours, therefore the estimated $p(b\vert b_{prev})$ has a wider confidence interval for the LLOY price series, as there are a smaller number of recorded bounce events.

In the price series with shuffled returns, there is no visually observable correlation between an SR level's bounce probability and its number of previous bounces. The estimated $p(b\vert b_{prev})$ are close to 0.5 across all investigated $b_{prev}$, with exceptions of higher $b_{prev}$ values. This unexpected deviation however can be explained by a smaller sample size, as is also indicated by a larger standard deviation. A close to 0.5 $p(b\vert b_{prev})$ is theoretically in agreement with martingale models, and suggests that this class of models may not be adequate for capturing this effect in the actual price series.

\subsection{Statistical Testing}
\cite{memory} have proposed the use of a Kolmogorov-Smirnov test to show that the estimated $p(b\vert b_{prev})$ for the original and the shuffled returns price series are from different distributions \citep{memory}. They have also conducted a chi-squared test of independence for the estimated $p(b\vert b_{prev})$ for $b_{prev}\in\{1, 2, 3, 4\}$, and concluded that $p(b\vert b_{prev})$ is statistically dependent on $b_{prev}$ for the original price series. For the purpose of showing the price series is predictable, which essentially contradicts the efficient market hypothesis, statistically significant results of deviation from the baseline of $p(b\vert b_{prev})=0.5$ is sufficient. To this end, we would like to show that the estimated $p(b\vert b_{prev})$ is higher for the original price series than the shuffled returns price series. A modified permutation test is used to estimate the expected probability of a higher bounce probability in the original than the shuffled returns series. We define this expectation to be $\Lambda$ and it can be written as
\begin{equation}\label{lambda equation}
\Lambda=\mathbb{E}\Big\{p\big[p_o(b\vert b_{prev})>p_s(b\vert b_{prev})\big]\Big\},
\end{equation}
where $p_o(b\vert b_{prev})$ and $p_s(b\vert b_{prev})$ refer to the probability of bouncing in the original and the shuffled returns price series respectively. This expectation can be numerically estimated using Monte Carlo simulations by making the $p_o(b\vert b_{prev})>p_s(b\vert b_{prev})$ comparison using a large number of price series with differently shuffled return. The procedures can be repeated for other lag window lengths, assets and support/resistance/combined levels.

\begin{figure}[!htbp]
\begin{center}
\includegraphics[width=13cm]{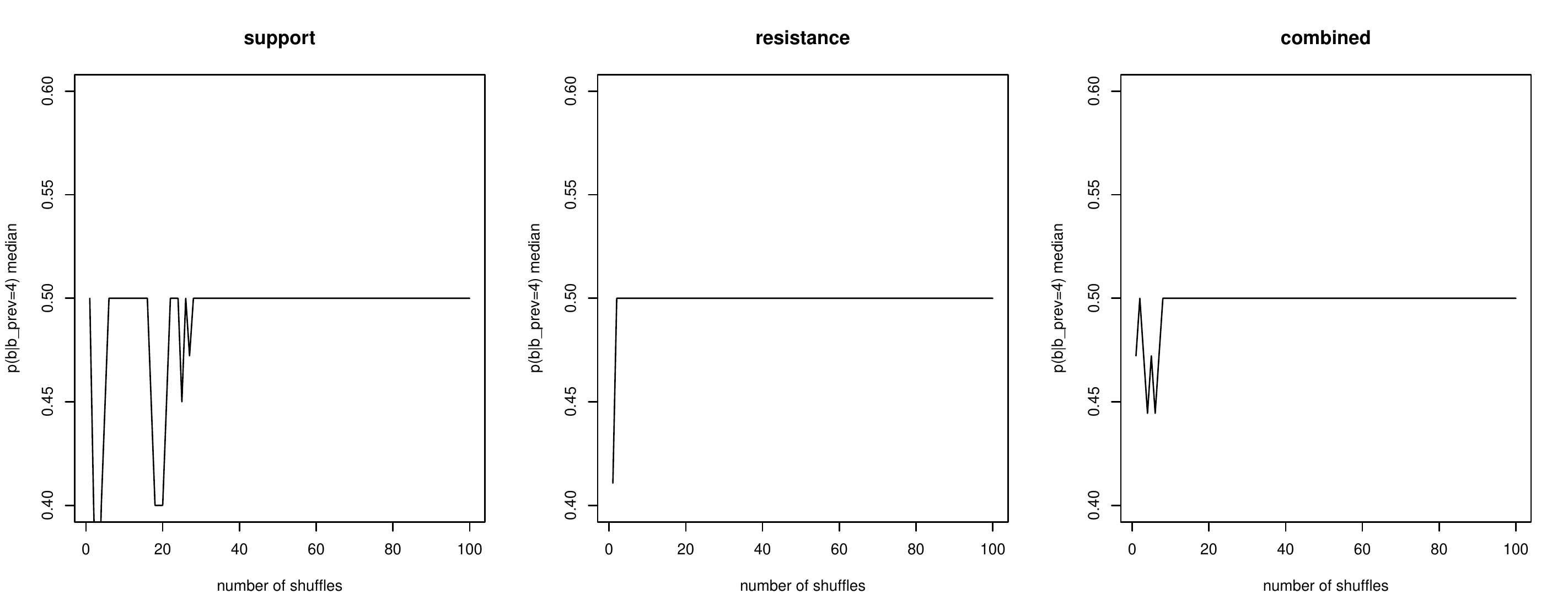}
\end{center}
\caption{Aggregated Median of Estimated $p_s(b\vert b_{prev}=8)$ for LLOY}\label{shuffles justification}
\end{figure}

A down side of this approach is that it is computationally expensive to apply the discovery and evaluation algorithms in Section \ref{algo section} to a large number of price series. Therefore it is of interest to use as few shuffled returns series as possible, as long as it is justified to do so. From the previous section we see that the number of recorded SR entry events for $b_{prev}=8$ can be very low for a shuffled returns series. Therefore perhaps the main concern of only using a small number of shuffled returns series is the instability of the estimation of $p_s(b\vert b_{prev}=8)$. The worst case for our investigation would be the estimation of $p_s(b\vert b_{prev}=8)$ for the LLOY price series for a lag window of 240 minutes. This is because the price series is the shortest, and leads to a very low number of events for estimation when coupled this with a long lag window. However, we find that the median of the estimated $p_s(b\vert b_{prev}=8)$ for LLOY stabilises after aggregating the results of 30 or more shuffled returns series. The median also converges to approximately 0.5, which is an expected result and confirms our random walk assumption. Figure \ref{shuffles justification} shows the evolution of this median over the number of shuffled returns series used for LLOY with a lag window of 240 minutes. In addition, since we would also like sufficient precision for significance level testing, we decide to evaluate 1000 shuffled returns series.

Table \ref{permutation test results} displays the estimated $\Lambda$ (Equation \ref{lambda equation}) for support, resistance and combined level, for $b_{prev}\in\{1, 2, 3, 4, 5, 6, 7, 8\}$. The large majority of the estimated $\Lambda$ are larger than 0.95, providing statistical evidence that the price series do have higher probability of bouncing on an SR level than their shuffled returns counterparts; the implication of this conclusion is that there is indeed a memory effect in the price series. 
However, due to LLOY having a shorter price series, a comparatively smaller number of SR level entry events with a high value of $b_{prev}$ is discovered from its shuffled returns series. This leads to very small $\Lambda$ estimates for the LLOY price series with 0.226 being the smallest. We are therefore apprehensive about making the same conclusion for the SR levels in LLOY for $b_{prev}$ values of 6, 7 and 8. Given the statistical evidence, we conclude SR levels have a higher probability of SR level bouncing in the LLOY price series than in the shuffled returns series up to and including $b_{prev}=5$.

\begin{table}[!htbp]
\begin{center}

\footnotesize
\setlength{\tabcolsep}{4pt}
\begin{tabular}{ccccccccccc}
\hline
&&&\multicolumn{8}{c}{$b_{prev}$}\\
asset & type & lag (minutes) &1&2&3&4&5&6&7&8\\\hline
EURUSD & Support & 60&1.000&1.000&1.000&1.000&1.000&1.000&0.942&0.991\\
EURUSD & Resistance & 60&1.000&1.000&1.000&1.000&1.000&1.000&0.995&0.893\\
EURUSD & Combined & 60&1.000&1.000&1.000&1.000&1.000&1.000&0.998&0.986\\
EURUSD & Support & 240&1.000&0.970&0.957&1.000&1.000&0.998&0.976&0.908\\
EURUSD & Resistance & 240&1.000&1.000&1.000&1.000&1.000&0.996&0.975&0.955\\
EURUSD & Combined & 240&1.000&1.000&1.000&1.000&1.000&1.000&0.998&0.978\\
LLOY & Support & 60&1.000&1.000&1.000&1.000&1.000&0.978&0.944&0.999\\
LLOY & Resistance & 60&1.000&1.000&1.000&1.000&1.000&0.931&0.412&0.226\\
LLOY & Combined&60&1.000&1.000&1.000&1.000&1.000&0.992&0.889&0.955\\
LLOY & Support & 240&1.000&1.000&0.998&0.999&0.997&0.740&0.916&0.822\\
LLOY & Resistance & 240&0.995&1.000&1.000&0.983&0.933&0.663&0.917&0.296\\
LLOY & Combined&240&1.000&1.000&1.000&1.000&0.998&0.753&0.972&0.614\\
BRENT & Support & 60&1.000&1.000&1.000&1.000&1.000&1.000&0.993&0.933\\
BRENT & Resistance & 60&1.000&1.000&1.000&1.000&1.000&0.998&0.865&0.902\\
BRENT & Combined & 60&1.000&1.000&1.000&1.000&1.000&1.000&0.996&0.946\\
BRENT & Support & 240&1.000&1.000&1.000&1.000&0.998&0.992&0.946&0.935\\
BRENT & Resistance &240&1.000&1.000&1.000&1.000&0.992&0.938&0.898&0.999\\
BRENT & Combined & 240&1.000&1.000&1.000&1.000&1.000&1.000&0.978&0.962\\\hline

\end{tabular}

\caption{Permutation Test Estimated $\Lambda$}\label{permutation test results}
\end{center}
\end{table}

\section{Temporal Decay of SR Levels}\label{temporal section}
Results in the previous section show a smaller bounce probability when applying the discovery algorithm using a 240 minute lag window, instead of a 60 minute lag window for the original series. So far our analysis which corroborate the work of Garzarreli  et  al. \citep{memory}, have not investigated the temporary nature of SR levels (though Osler \citep{osler} has concluded that SR levels can last up to at least 5 business days). In this section, we extend the knowledge of SR levels in the literature by considering the temporal aspect of SR levels on an intraday basis. The duration of which individual SR levels exist are non-deterministic and have varying lengths. We therefore investigates if and how the memory effect decays over time by aggregating results over all discovered SR levels. We obtain evidence for the decay in memory effect in terms of bounce probability in the macro scale by varying lag window lengths, and in the micro scale by measuring time from last bounce. The decay behaviour seems to be unique depending on the particular asset, which we attribute to the general price trend of the asset in 2018 and to the unique characteristic of the asset.

\subsection{Macro Decay}
\begin{figure}[!htbp]
\begin{center}
\includegraphics[width=13cm]{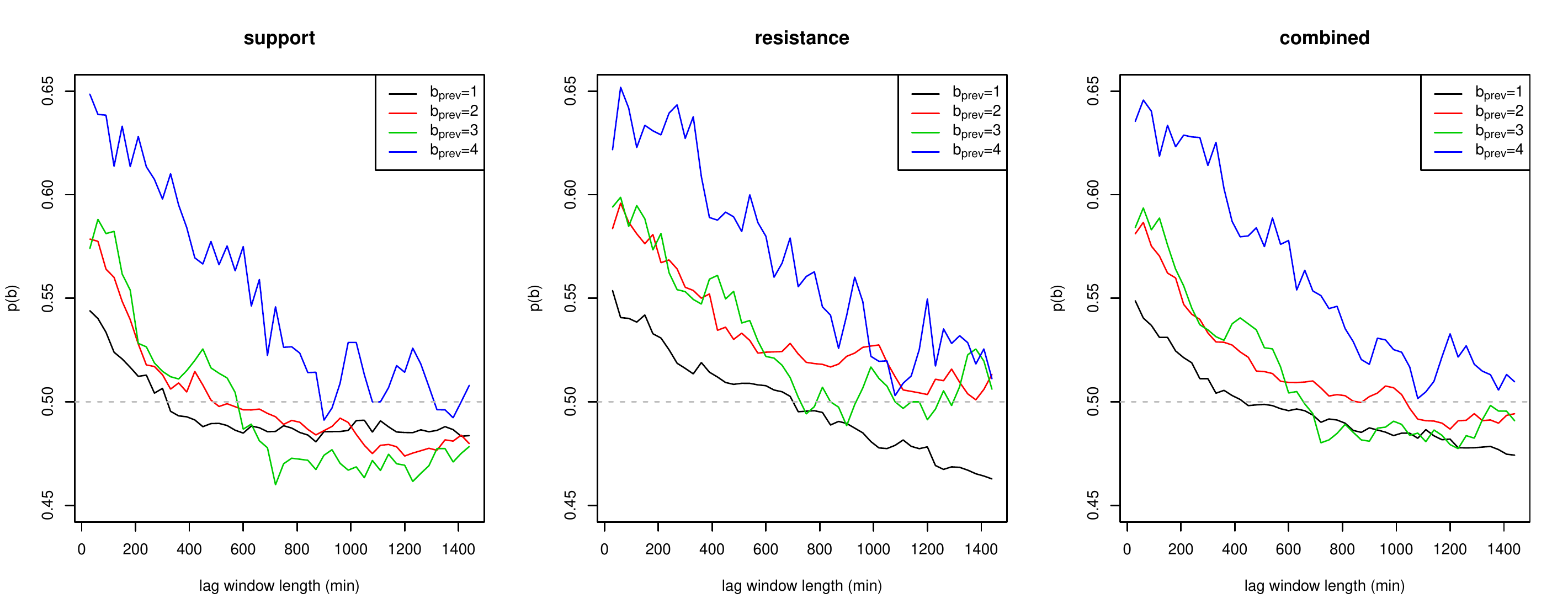}
\end{center}
\caption{\textbf{EURUSD Lag Window Length Analysis.}}\label{eurusd lag analysis}
\end{figure}

\begin{figure}[!htbp]
\begin{center}
\includegraphics[width=13cm]{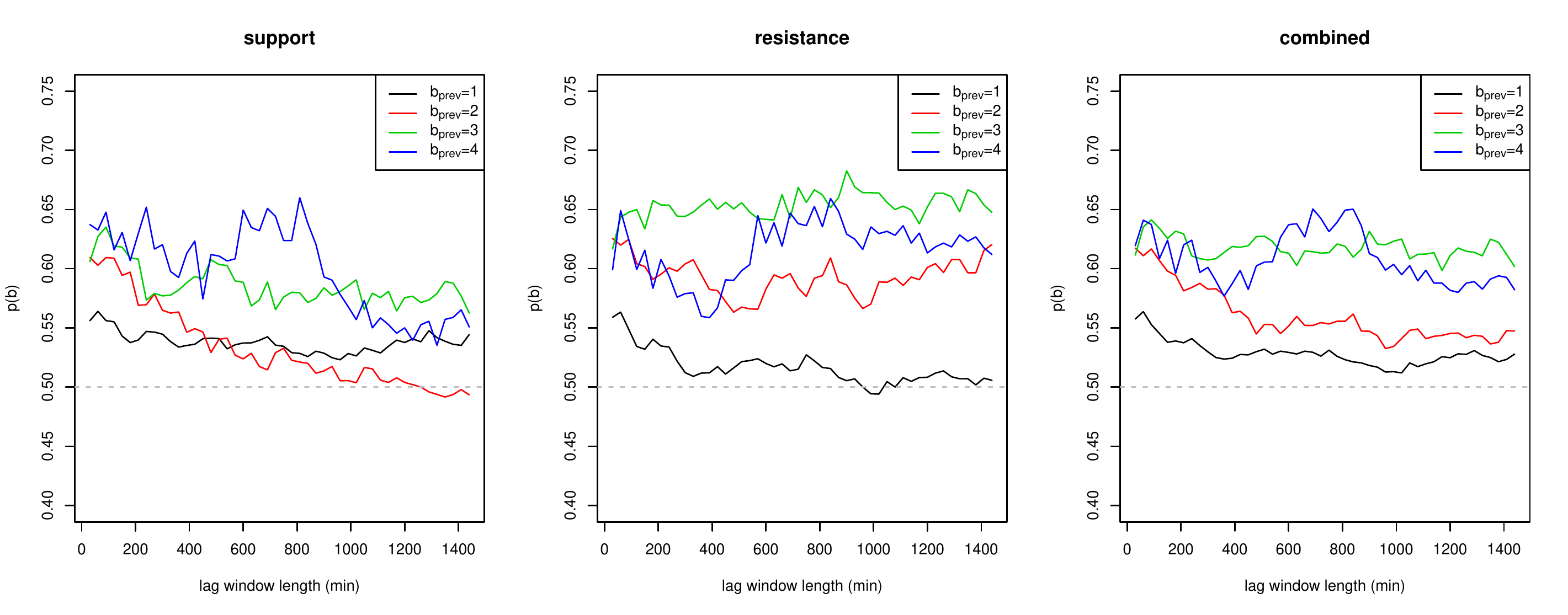}
\end{center}
\caption{\textbf{LLOY Lag Window Length Analysis.}}\label{lloy lag analysis}
\end{figure}

\begin{figure}[!htbp]
\begin{center}
\includegraphics[width=13cm]{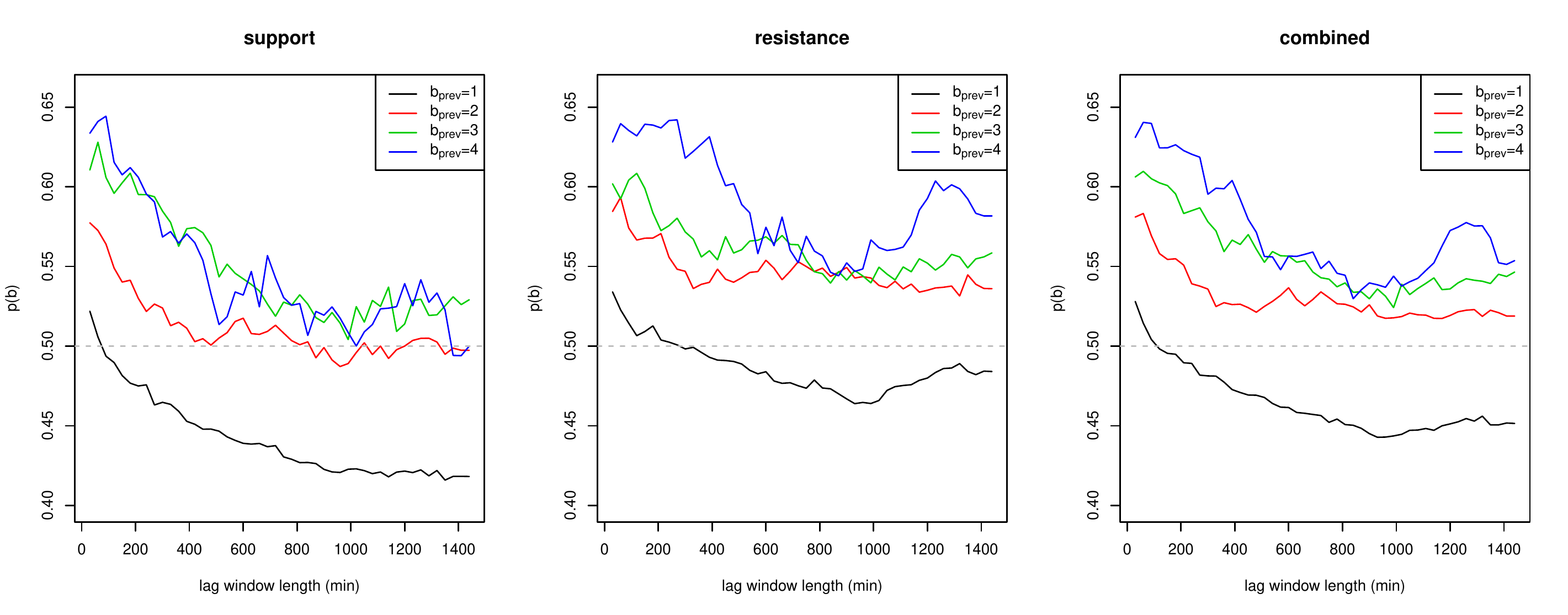}
\end{center}
\caption{\textbf{BRENT Lag Window Length Analysis.}}\label{brent lag analysis}
\end{figure}

We refer the macro decay in the SR level memory effect as the negative correlation observed for SR level bounce probabilities and the lag window length used for the discovery of the SR levels. To observe this decay directly, we compare the estimated $p(b\vert b_{prev})$ for lag windows from 30 minutes to 1440 minutes (1 day) in 30 minutes increments. It is anticipated that the use of a large lag window such as 1440 minutes gives a small number of discovered SR levels with a higher number of previous bounces, which in turn leads to inaccurate estimations. To overcome this, we focus our attention only on SR levels with 1 to 4 previous bounces. Figure \ref{eurusd lag analysis}, \ref{lloy lag analysis} and \ref{brent lag analysis} display estimated $p(b\vert b_{prev})$ over different lag window lengths and overlay $b_{prev}$ values from 1 to 4 for comparison. 

As expected a general downward trend for the estimated $p(b\vert b_{prev})$ against lag window length is observed in Figure \ref{eurusd lag analysis}. This shows that previous bounces of an SR level over a longer time frame become less effective at predicting future bounces. For EURUSD, previous bounces of an SR level predicts future bounces most effectively with short lag windows, but we also observe that SR levels with a higher number of previous bounces decay in predictive ability the slowest. For example, the estimated bounce probability for SR levels with 1 previous bounce decays to 0.5 at approximately a 350 minute lag window length, whereas SR levels with 4 previous decays to the same level at a much longer 900 minute lag window length. The comparatively quicker decay of support levels to a bounce probability of less than 0.5 can be attributed to a general bear market in 2018 for EURUSD.

The decay behaviour of the LLOY asset is very different than that of EURUSD. Figure \ref{lloy lag analysis} shows that most bounce probabilities do not decay to 0.5 or below. This suggests the memory effect of SR levels are more persistent in the stock markets. In fact, it is difficult to determine, from inspecting Figure \ref{lloy lag analysis} alone, if the memory effect of SR levels decays at all for the LLOY asset. Again, a somewhat observable decay in bounce probability for the support level compared to the resistance level can be attributed to the general bear market for LLOY in 2018. A general decay in bounce probability for longer lag window lengths is observed for BRENT in Figure \ref{brent lag analysis}, which is similar to that of EURUSD. However, it would seem the memory effect for the SR levels with only 1 previous bounce decays very rapidly to 0.5 at lag window lengths of no longer than 300 minutes; it also decays to lower than 0.5 which suggests not only does 1 previous bounce not predict future bounces, it actually predicts penetrations. This could be a characteristic of the BRENT asset, but can also be attributed to an extreme bear market in the final quarter of 2018 for the BRENT asset, as this feature is more prominent for support levels than it is for resistance levels.

\subsection{Micro Decay}
Next we investigate the decay in bounce probability as time increases from the previous bounce of an SR level, referred as micro decay in this section. To this end, we measure the time from previous bounce of an SR level for every SR level entry event and construct logistic regression models to describe the relationship between price bounces and time from previous bounce. This is done separately for $b_{prev}$ values from 1 to 8, and for a lag window length of 600 minutes. We have picked this particular length of lag window (600 minutes) because most estimated bounce probability manages to decay to 0.5 in the macro scale in the 600 minute lag window as described in the previous section, and therefore allows us to study the micro decay. On the contrary, if we have picked a 60 minute lag window length, no micro decay can be observed as we have already shown there is minimal macro decay within a 60 minute time frame. The logistic model assumes the following form:
\begin{equation*}
\log\Big(\frac{Y}{1-Y}\Big)=a+bX,
\end{equation*}
where $Y$ represents the binary variable for price exiting SR level events (1 for bounce events, 0 for penetration events), $X$ the time from previous bounce and $a, b$ the model parameters. The inclusion of the intercept parameter $a$ serves as a baseline bounce probability, but we are most interested in the $b$ parameter which indicates the relationship between bounce probability and time from previous bounce of an SR level. Non-linear transformations of $X$ have also been modelled but they do not improve model fit.

The estimated regression parameters for the EURUSD price series are displayed in Table \ref{eurusd regression table}. It can be observed that not all estimated paramters are statistically significant. However, the signs of the estimated parameters are consistent with previous results and our expectations. The estimated values for $a$ are positive, implying a baseline bounce probability of higher than 0.5; this number increases with $b_{prev}$ which is consistent with our previous conclusion of multiple SR level bounces confirms its strength. The estimated $b$ values are negative, implying a negative correlation with bounce probability and time from previous bounce; confirming our suspicion of a micro decay in bounce probability. Despite both of these observations, the inference of the existence of a micro decay is only supported by models for $b_{prev}=4, 5$ in which both estimated parameters are statistically significant. The comparatively smaller sample size for $b_{prev}=6, 7, 8$ regression models contributes to the lack of statistically significant parameter estimates, and this also holds true for the LLOY and BRENT price series.

\begin{table}[!htbp]
\centering
\setlength{\tabcolsep}{20pt}
\begin{tabular}{cccc}
\hline
$b_{prev}$ & $a$ & $b$ & $N$ \\\hline
1 & 0.00481 & -0.00052 & 3391\\
2 & 0.07202 & -0.00075 & 1973\\
3 & 0.05986 & -0.00097 & 1051\\
4 & 0.40241*** & -0.00179* & 505\\
5 & 0.40331** & -0.00273* & 258\\
6 & 0.64003*** & -0.00187 & 151\\
7 & 0.45755 & -0.00300 & 89\\
8 & 0.71684* & -0.00158 & 41\\
   \hline
\end{tabular}
\caption{\textbf{EURUSD Micro Decay Logistic Regression Estimated Parameters (600 min lag window).} P-values are denoted: $<$0.001'***', $<$0.01'**' and $<$0.05'*', $N$ denotes the regression sample size.}\label{eurusd regression table}
\end{table}

The estimated regression parameters for the LLOY price series are displayed in Table \ref{lloy regression table}. Similar to the result for the EURUSD price series, all estimated $a$ values are positive and all estimated $b$ values are negative, which are consistent with previous results and within our expectations. Although the macro decay in bounce probability is conservatively speaking inconclusive as seen in Figure \ref{lloy lag analysis}, the negative estimated $b$ values suggest the existence of a micro decay. There is no statistically significant evidence to support this suggestion, but the memory effect in the LLOY price series seems to be more persistence which  explains this result.

\begin{table}[!htbp]
\centering
\setlength{\tabcolsep}{20pt}
\begin{tabular}{cccc}
\hline
$b_{prev}$ & $a$ & $b$ & $N$ \\\hline
1 & 0.15099* & -0.00066 & 1024\\
2 & 0.23152** & -0.00048 & 667\\
3 & 0.56173*** & -0.00236 & 378\\
4 & 0.62520*** & -0.00143 & 224\\
5 & 0.74737*** & -0.00266 & 117\\
6 & 0.38472 & -0.00670 & 69\\
7 & 1.13922 & -0.00221 & 39\\
8 & 0.45460 & -0.00574 & 24\\
   \hline
\end{tabular}
\caption{\textbf{LLOY Micro Decay Logistic Regression Estimated Parameters (600 min lag window).} P-values are denoted: $<$0.001'***', $<$0.01'**' and $<$0.05'*', $N$ denotes the regression sample size.}\label{lloy regression table}
\end{table}

The estimated regression parameters for the BRENT price series are displayed in Table \ref{brent regression table}. All estimated $a$ values are positive except for the $b_{prev}=1$ model, but this is easily explained by the significant macro decay observed in Figure \ref{brent lag analysis}. The estimated $b$ values do not have a consistent sign which makes it difficult to say if micro decay exists. However, the model for $b_{prev}=4$ has both estimated parameters to be statistically significant, which implies a higher than 0.5 bounce probability baseline and the existence of a micro decay.

\begin{table}[!htbp]
\centering
\setlength{\tabcolsep}{20pt}
\begin{tabular}{cccc}
\hline
$b_{prev}$ & $a$ & $b$ & $N$ \\\hline
1 & -0.14466*** & -0.00024 & 2785\\
2 & 0.14510** & 0.00004 & 1500\\
3 & 0.28176*** & -0.00118 & 839\\
4 & 0.00044*** & -0.00195** & 467\\
5 & 0.00982** & -0.00049 & 239\\
6 & 0.32726 & 0.00146 & 144\\
7 & 0.17473 & 0.00093 & 76\\
8 & 0.32612 & 0.00347 & 39\\
   \hline
\end{tabular}
\caption{\textbf{BRENT Micro Decay Logistic Regression Estimated Parameters (600 min lag window).} P-values are denoted: $<$0.001'***', $<$0.01'**' and $<$0.05'*', $N$ denotes the regression sample size.}\label{brent regression table}
\end{table}

\section{SR Levels in AR(1) Processes}\label{AR section}
\begin{figure}[!htbp]
\begin{center}
\includegraphics[width=13cm]{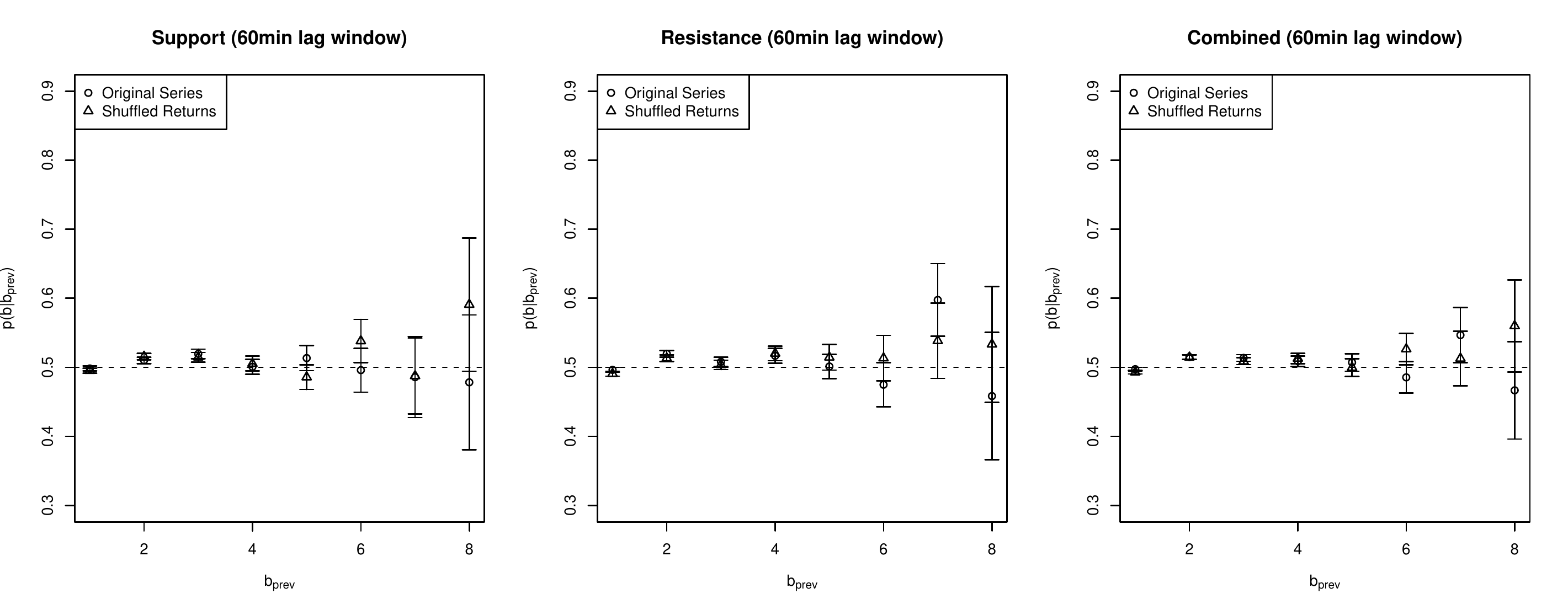}
\end{center}
\caption{\textbf{Estimated $p(b\vert b_{prev})$ for AR1 Process with $\rho=1$.} The confidence intervals are constructed using a width of 1 standard deviation on each side. }\label{rho1}
\end{figure}

\begin{figure}[!htbp]
\begin{center}
\includegraphics[width=13cm]{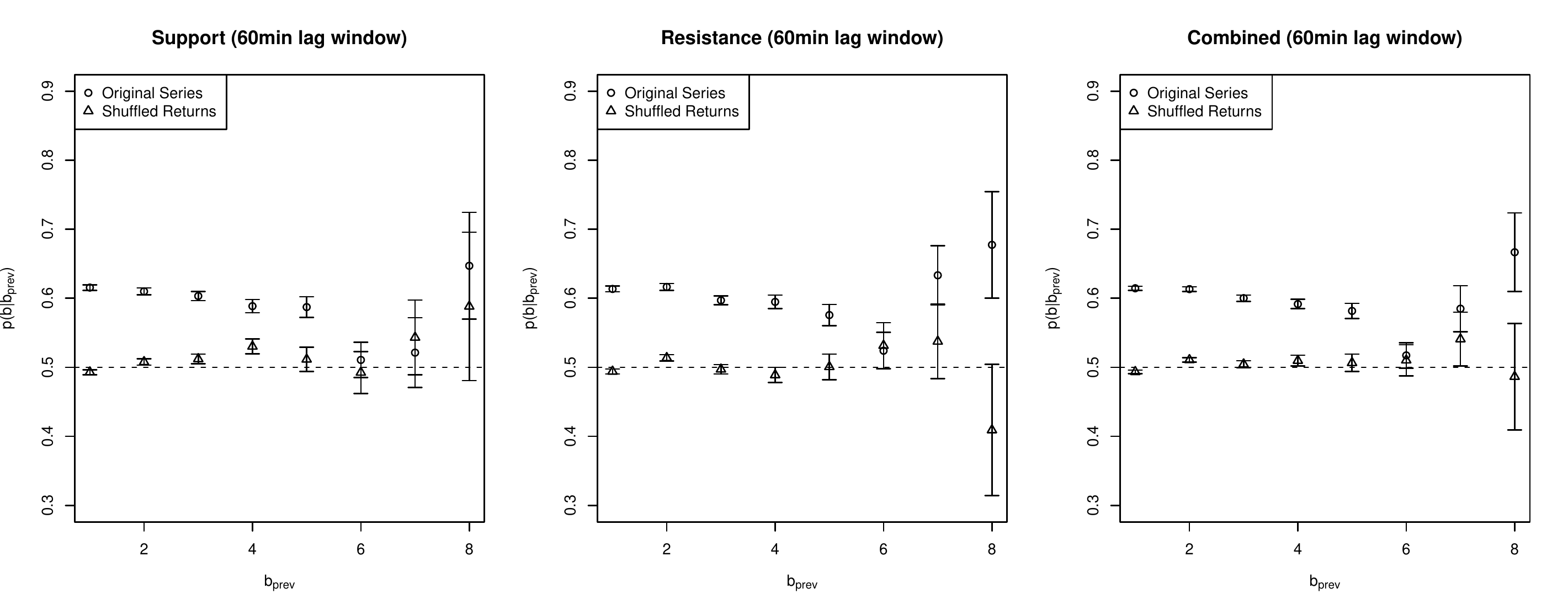}
\end{center}
\caption{\textbf{Estimated $p(b\vert b_{prev})$ for AR1 Process with $\rho=0.95$.} The confidence intervals are constructed using a width of 1 standard deviation on each side.}\label{rho2}
\end{figure}

\begin{figure}[!htbp]
\begin{center}
\includegraphics[width=13cm]{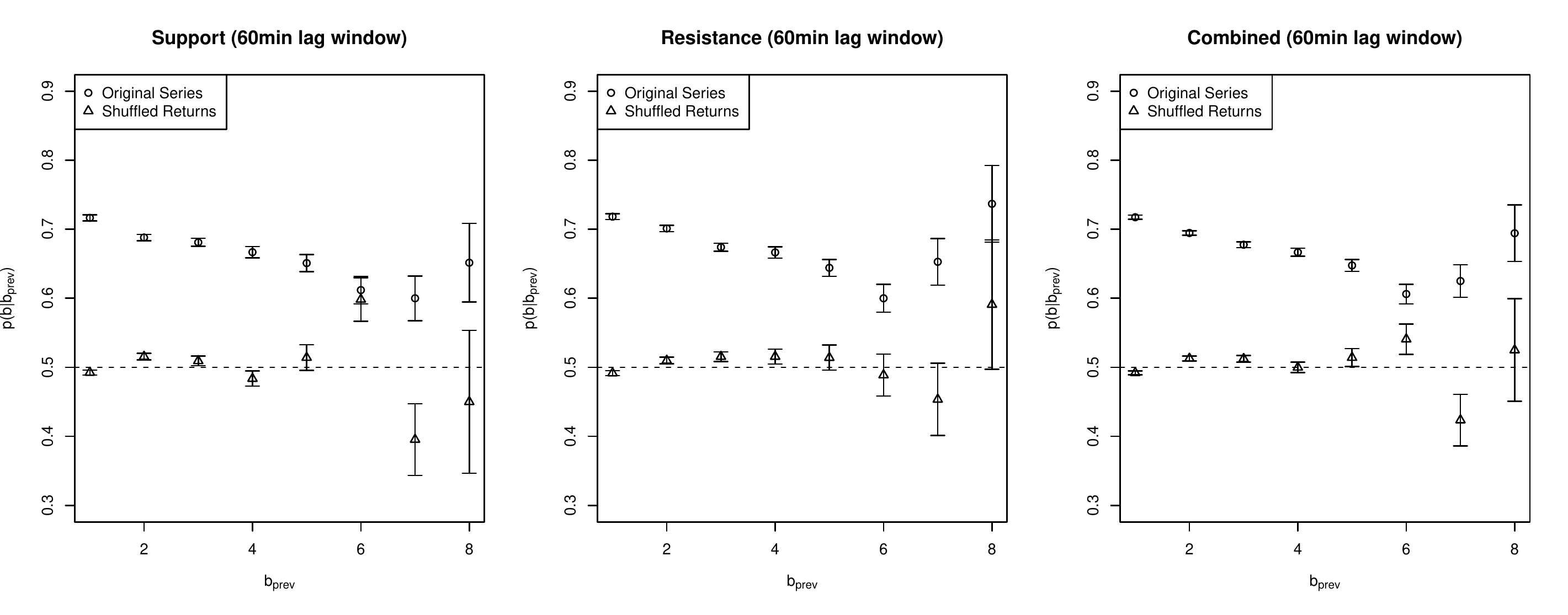}
\end{center}
\caption{\textbf{Estimated $p(b\vert b_{prev})$ for AR1 Process with $\rho=0.9$.} The confidence intervals are constructed using a width of 1 standard deviation on each side.}\label{rho3}
\end{figure}

A simplified interpretation of the efficient market hypothesis is that financial time series should follow a random walk (or Brownian motion in continuous time) like pattern. This feature implies future prices cannot be predicted using historical data. Auto-regressive processes with unit roots simulate this feature, and thus studying SR levels in simulated auto-regressive processes allows us to determine if SR levels naturally exist in efficient markets. We investigate this line of argument in this section by modelling the price series with auto-regressive processes with 1 lag term (AR1 processes), and then applying the discovery and evaluation algorithms in Section \ref{algo section} on the simulated time series with various extend of stationarity to determine if results obtained from the previous section can be replicated with simple mean reverting time series. Simulated time series has the following form:
\begin{equation*}
X_t=\rho X_{t-1}+\epsilon_t,
\end{equation*}
where $\epsilon_t\sim N(0, 1)$. The parameter $\rho$ in this model determines the strength of stationarity of the time series.  
We simulate time series for different values of $\rho$, each with a length of 1,000,000; and then apply the discovery and evaluation algorithms in Section \ref{algo section} with a 60 time unit lag window. Figure \ref{rho1}, \ref{rho2} and \ref{rho3} displays the estimated $p(b\vert b_{prev})$ for AR1 processes with $\rho=1$, $\rho=0.95$ and $\rho=0.9$ respectively.

For simulated AR1 series with $\rho=1$, there is no visually discernible difference for the estimated $p(b\vert b_{prev})$ between the original series and shuffled returns series. This means that if the time series is a random walk, there is no memory in the time series and that the probability of bouncing again of a discovered SR level is not dependent on the number of previous bounces. This also shows that a time series with shuffled returns can be interpreted as a random walk for our purposes. However, if we simulate stationary AR1 series, such as that of $\rho=0.95$ and $\rho=0.9$, we can see that smaller $\rho$ increases the values of the estimated $p(b\vert b_{prev})$. This is within our expectation as we would expect stronger stationarity, in terms of the parameter $\rho$, to increase the probability of mean reversion, regardless of the existence of SR levels. 

The results displayed here differs from the EURUSD, LLOY and BRENT price series in that $b_{prev}$ values seem to have a negative correlation with the probability of bouncing. An explanation for this would be that it is statistically less likely to have higher bounces on a recent maximum/minimum, if a genuine SR level does not exist. Independence of previous bounces means that the probability of bouncing can be interpreted as independent Bernoulli events with $p(b)$, and that decreasing the value of $\rho$ simply increases the value of $p(b)$; therefore having multiple bounces on the same level in a lag window, is akin to having a bounce event happening multiple times, which yields a smaller probability according to the multiplicative rule of probability, assuming $p(b)<1$. We can therefore conclude that the results we obtain from the financial price series cannot be replicated from simple stationary AR1 processes; and some that other features, which very well could be SR levels, are in play to create price trend reversals at recent price maximum/minimum.

\section{Conclusion \& Future Work}\label{conclusion section}
In this paper, the existence of SR levels and the tendency of price trends reversing at these levels are investigated. Stacking of limit orders and trader psychology are two possible, though not mutually exclusive, reasons for the existence of SR levels. Though qualitative knowledge of SR levels has been accumulating for at least 3 decades in the retail trading circle, no formal quantitative definition has been established. We therefore propose a general definition of SR levels suitable for the evaluation framework used throughout this paper.

Focusing on intraday short term price trend reversals, we measure the ability of SR levels to reverse price trends by the probability of price bouncing on SR levels within a short intraday time frame. Using a heuristic method of discovering SR levels, we gathered empirical evidence that price have a higher probability of bouncing on discovered SR levels. We also find that the number of previous bounces of an SR level positively correlates with price bouncing on it again. This result is statistically significantly different from the same price series with shuffled returns, exhibits temporal decay which is dependent on the particular asset and cannot be replicated simply by simulated AR(1) processes, stationary or otherwise.

Moving forward, we would like to analyse limit order book and trade data in order to improve the current SR level discovery methodology in relation to the market micro-structure. Using the discoveries from this analysis, we can then attempt to model SR levels with agent based models to simulate behaviour and consequences of market participants. Finally we would like to explore methods to incorporate SR levels into algorithmic trading strategies such as optimal liquidation/acquisition in continuous finance.

\section*{Acknowledgements}
The authors thank the UK PhD Centre in Financial Computing at University College London for their financial support for Ken Chung.

\end{document}